\documentclass{article}

\usepackage{amssymb}
\usepackage{amsmath}
\usepackage{graphicx}
\usepackage[latin1]{inputenc}
\usepackage{epsfig}
\usepackage{float}
\usepackage{subfigure}
\usepackage{psfrag}
\usepackage[T1]{fontenc}
\usepackage{color}
\usepackage{array}
\usepackage{dsfont}
\usepackage{algorithm,algorithmic}
\usepackage{authblk}
\usepackage[bookmarks=false,hidelinks]{hyperref}
\usepackage{listings}
\usepackage{a4wide}
\usepackage{tikz}
\usetikzlibrary{arrows,decorations.pathmorphing,backgrounds,fit,positioning,shapes.symbols,chains}
\definecolor{greenMatlab}{RGB}{46,139,87} 
\newcommand{\blue}[1]{\textcolor{blue}{#1}}

\newcommand{\charles}[1]{\blue{#1}}

\providecommand{\blue}[1]{\textcolor[RGB]{16,97,169}{#1}} 
\providecommand{\ie}[0]{\emph{i.e.} }
\providecommand{\eg}[0]{\emph{e.g.} }

\definecolor{bleuONERA}{RGB}{16,97,169}
\definecolor{grisONERA}{RGB}{64,64,66}

\providecommand{\bu}{\mathbf{u}} %
\providecommand{\by}[0]{\mathbf{y}} %
\providecommand{\Htran}[0]{\mathbf{H}} %
\providecommand{\Cplx}[0]{\mathbb{C}} %
\providecommand{\Real}[0]{\mathbb{R}} %
\newenvironment{eq}{\everymath {\displaystyle \everymath{ }} \equation}{ \endequation} %

\newtheorem{remark}{Remark}

\begin{document}

\title{Flow separation control design with experimental validation}

\author[1]{T. Arnoult}
\author[2]{G. Acher}
\author[2]{V. Nowinski}
\author[3]{P. Vuillemin}
\author[4]{C. Briat}
\author[1]{P. Pernod}
\author[1]{C. Ghouila-Houri}
\author[2]{A. Talbi}
\author[1]{E. Garnier}
\author[3]{C. Poussot-Vassal}
\affil[1]{Univ. Lille, CNRS, Centrale Lille, Univ. Polytechnique Hauts-de-France, UMR 8520 - IEMN - Institut d'Electronique de Micro\'electronique et de Nanotechnologie, F-59000 Lille, France.}
\affil[2]{Univ. Lille, CNRS, ONERA, Arts et Metiers Institute of Technology, Centrale Lille, UMR 9014, Laboratoire de M\'ecanique des fluides de Lille-Kamp\'e de F\'eriet, F-59000 Lille, France.}
\affil[3]{ONERA - The French Aerospace Lab, F-31055 Toulouse, France.}
\affil[4]{D-BSSE, ETH Z\"urich, Switzerland.}

\maketitle

\begin{abstract}
Flow control aims at modifying a natural flow state to reach an other flow state considered as advantageous. In this paper, active feedback flow separation control is investigated with two different closed-loop control strategies, involving a reference signal tracking architecture. Firstly, a data-driven control law, leading to a linear (integral) controller is employed. Secondly, a phenomenological/model-driven approach, leading to a non-linear positive (integral) control strategy is investigated. While the former benefits of a tuning  simplicity, the latter prevents undesirable effects and formally guarantees closed-loop stability. Both control approaches were validated through wind tunnel experiments of flow separation over a movable NACA 4412 plain flap. These control laws were designed with respect to hot film measurements, performed over the flap for different deflection angles. Both control approaches proved efficient in avoiding flow separation. 
The main contribution of this work is to provide practitioners simple but yet efficient ways to design a flow separation controller. In addition, a complete validation campaign data-set is provided.
\end{abstract}



\section{Introduction}
\label{sec:intro}
\subsection{Forewords on flow separation objective}

Flow separation over an aircraft flap is characterized by a decrease in the lift coefficient, an increase in the drag coefficient and can occur during the critical take-off and landing phases. Most aircraft circumvent this issue using slotted flaps, which however add structural weight and complicate the maintenance. Therefore, one solution would be to simplify these structures into plain flaps with integrated flow control devices to avoid flow separation.

Flow control consists in modifying a flow general behavior with a space localized perturbation, in order to reach a flow configuration considered as favorable. In that sense, flow control can help reducing noise radiation, delaying the laminar-turbulent boundary layer transition or can help avoiding flow separation \cite{gad-el-hak_flow_2000}. Flow control methods can be categorized as passive or active methods. Passive methods do not require any external source of energy to act on the flow. One may mention vortex generators, which have been widely used to prevent flow separation on aircraft wings. Vortices generated by these devices help re-energizing the boundary layer and therefore prevent its separation. However, they act permanently on the flow, even at off-points design. In that sense, an active flow control method can be employed in a closed-loop strategy as considered in this study. The actuators command can therefore be adapted depending on the needs.

As discussed by Pastoor \textit{et al.}  \cite{pastoor_feedback_2008}, closed-loop flow control strategies have been applied to different canonical flow configurations such as the flow around a cylinder, the flow over backward facing steps, the flow over open-cavities or separated flows over airfoils. These flows are described by different dynamics and different actuating strategies may be required to implement their closed-loop control. Considering the control of flow separation over airfoils, one may aim at modifying the mean flow properties. Several studies have focused on this case with different control methods and objectives. One control method consists in turning the actuators when a threshold value measured by the sensors is exceeded. This triggering method was used in several studies. For instance, Packard and Bons \cite{packard_closed-loop_2012} and Rethmel \textit{et al.} \cite{rethmel_flow_2011} studied the flow separation control over wings based on NACA airfoils shape. In these studies, hot films sensors are placed on the models. The RMS (Root Mean Square) value of the hot films voltage is used to define the threshold value, according to which the flow is separated. In a similar way, Lombardi \textit{et al.} \cite{lombardi_closed-loop_2013}, Tewes \textit{et al.} \cite{tewes_feedback-controlled_2011} and Benard \textit{et al.} \cite{benard_benefits_2011} propose a triggering control method based on pressure measurements. In \cite{lombardi_closed-loop_2013} the triggering control criterion is based on unsteady pressure measurements performed on the model. A spectrum analysis of the pressure measurements is employed to determine the energy at a characteristic frequency indicating the onset of flow separation. If this energy value is overshot, then the plasma actuator placed at the wing leading edge is turned on. Regarding the studies \cite{tewes_feedback-controlled_2011} and \cite{benard_benefits_2011}, the triggering control methods are coupled with an hysteresis effect. Instead of reattaching a separated flow over the considered wings, both study maintain the flow attached and therefore reduce the energy required to control the flow. In \cite{tewes_feedback-controlled_2011}, the threshold value is defined relatively to the leading edge pressure coefficient value, while in \cite{benard_benefits_2011} the threshold value is set with respect to the pressure coefficient RMS computed at the leading edge. A second control methodology is a model free approach based on gradient methods such as extremum seeking and slope seeking. For instance, Benard \textit{et al.} focus on flow reattachment over NACA 0015 airfoil with a slope seeking algorithm, which aims at maximising the lift coefficient. In a similar way, Becker \textit{et al.} \cite{becker_adaptive_2007} tend to maximize the lift coefficient of a high-lift configuration composed of a NACA 4412 main airfoil associated with a NACA 4415 flap, firstly with a SISO (Single Input Single Output) extremum seeking scheme, then with a SISO slope seeking algorithm, finally extended to a MIMO (Multiple Input Multiple Output) slope seeking approach. This study was carried out based on the pressure coefficient derived from unsteady pressure measurements. A third control approach is based on the use of black box models identifying the system's input and output transfer. Recently, Sanchez \textit{et al.} \cite{Sanchez:2020} explored a sliding mode approach and applied it on a numerical example.  King \textit{et al.} \cite{king_flight_2013} consider in their study the control of flow separation on a wing of a 1:2 scale model proposed by Airbus and expand their experiments on a full scale glider. In both cases, they identify a model using PRBS (Pseudo-Random Binary Signals) to drive the pulsed jets. From the black box model, the $\mathcal{H}_\infty$ synthesis is employed to design a robust controller. Different type of controllers can be synthesized based on this black box model approach (see \eg \cite{lee_closed-loop_2013}).

\subsection{Contribution statement}

\begin{figure}[h]
    \centering
    \includegraphics[width=.8\columnwidth]{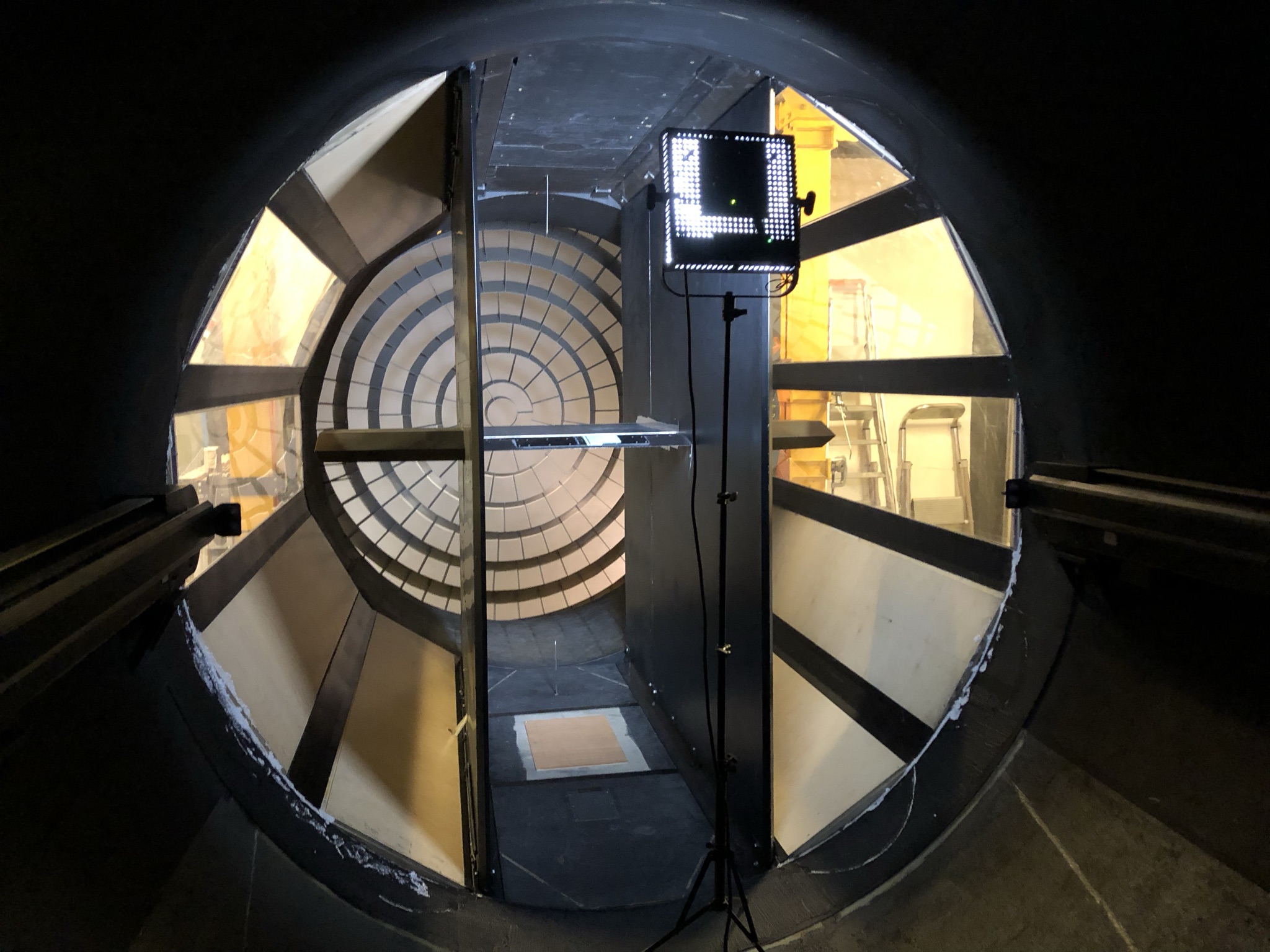}
    \caption{Wind tunnel facility view (Onera, Lille, France). The commanded horizontal wing is in between the two vertical structures. The flow is longitudinally travelling from the back of the photo.}
    \label{fig:setup}
\end{figure}

The contribution of the paper is to deploy and evaluate in an experimental wind tunnel facility, involving a NACA 4412 plain flap airfoil, two active closed-loop control design strategies to drive the flow separation phenomena. The first one is a (model-free) \emph{linear data-driven} approach, while the second one is a \emph{positive nonlinear phenomenological/model-driven} strategy. The data-driven rationale is extensively detailed in \cite{KergusPhD:2019}, while the positive strategy is discussed in \cite{BriatSIADS:2020}, initially considered for the control of biological systems in \cite{BriatCellSystems:2016}. Both control methods are based on the use of on/off solenoid valves as actuators and on hot film sensors. Both control structure involve the same reference signal to track, which value is also discussed. Each strategy is validated in a wind tunnel facility (see Figure \ref{fig:setup}) and leads to a lift increase, and cancelled / reduced flow separation. 

\begin{figure}[h]
    \centering
    \includegraphics[width=\columnwidth]{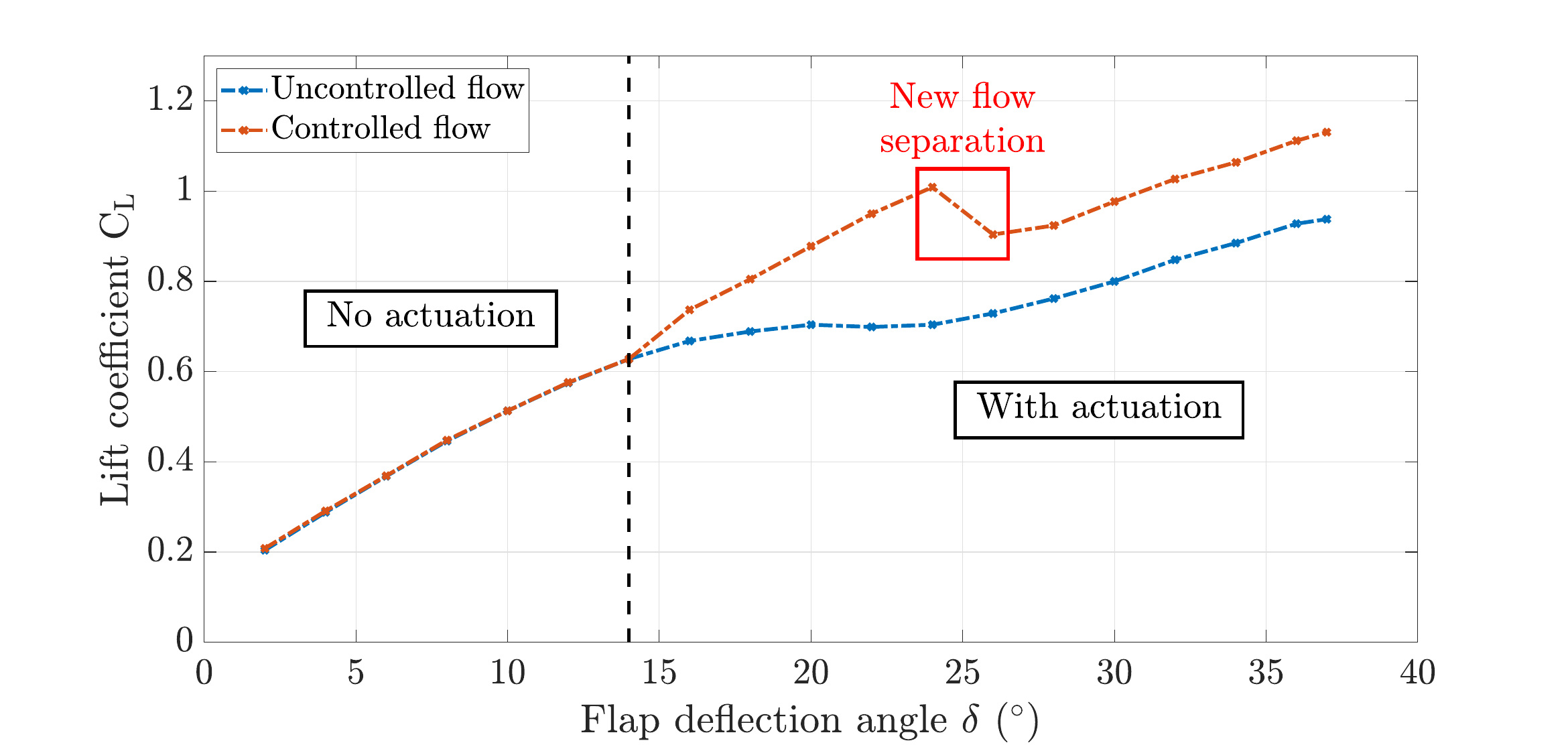}
    \caption{Evolution of the lift coefficient ${C_L}$ without (blue curve) and with (red curve) control against the flap deflection angle ${\delta}$ for ${U_\infty =\ 34.5}$ m/s.}
    \label{fig:performanceObjective}
\end{figure}

\begin{figure}[h]
    \centering
    \includegraphics[width=.8\columnwidth]{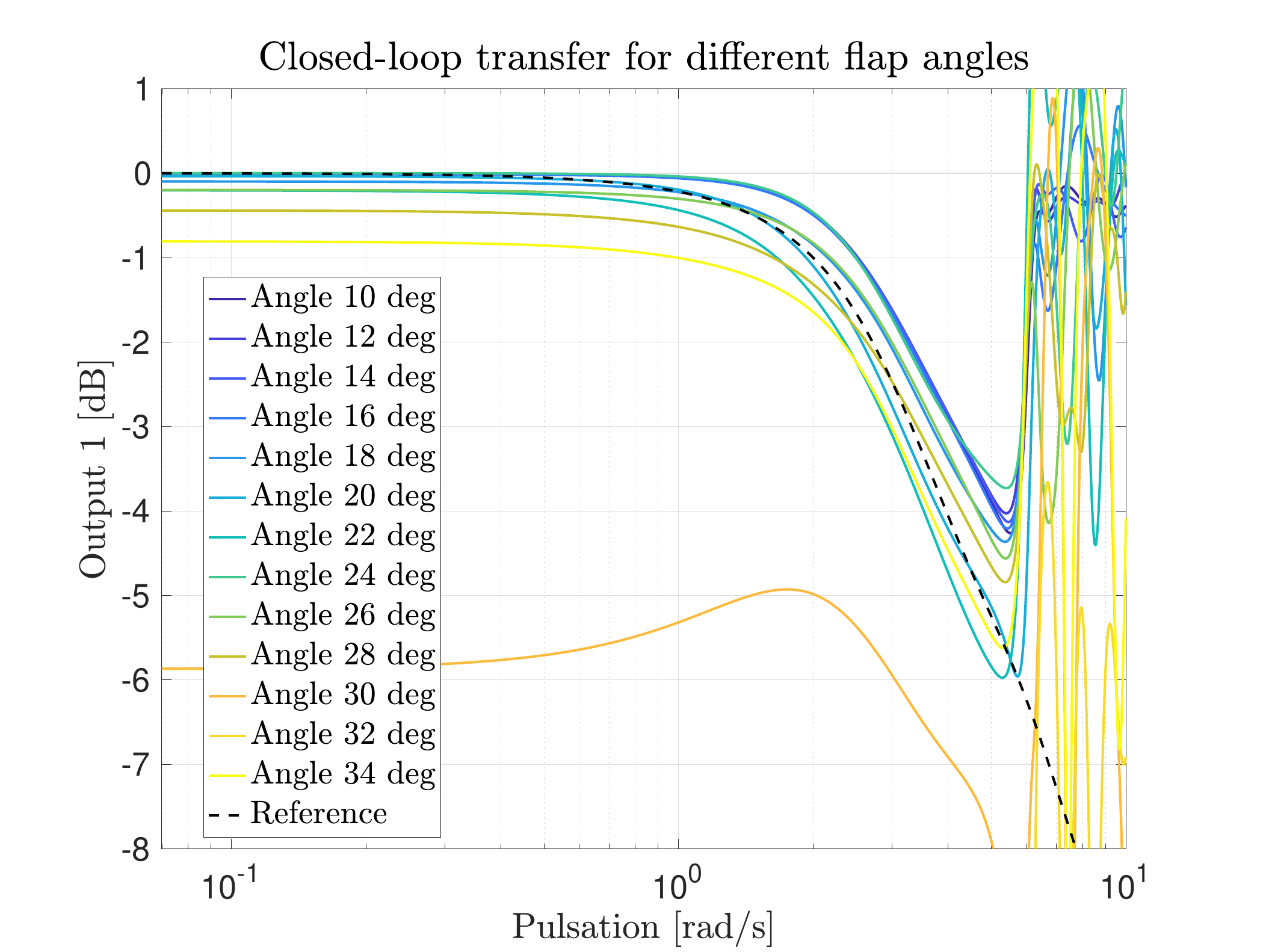}
    \caption{Frequency-domain responses of the controlled system for ${U_\infty =\ 34.5}$ m/s. Coloured solid lines (response for different flap deflection angle ${\delta}$) and reference (dashed black).}
    \label{fig:performanceObjectiveFR}
\end{figure}

As a glimpse of this paper result, Figure \ref{fig:performanceObjective} illustrates the lift performances with and without flow separation control. In addition, Figure \ref{fig:performanceObjectiveFR} illustrates the closed-loop frequency response (obtained with a frequency  sweeping reference signal) for varying flap angles, with respect to the reference objective transfer. The rest of this note details the process for reaching such performances and derives a generic but yet simple approach to design and validate two flow separation feedback control laws. We believe that the proposed rationales are sufficiently simple to be applied on a variety of similar setups.

\subsection{Notations and paper organisation}

After recalling the flow separation problem in section \ref{sec:intro}, section \ref{sec:setup} presents the considered experimental setup. Both linear data-driven and nonlinear positive model-driven flow control designs are recalled and validated in section \ref{sec:control}. Conclusions and outlooks are gathered in section \ref{sec:conclusion}.

By $\Real$, $\mathbb Z$ and $\mathbb N$ we indicate the sets of real, integer and natural (positive integer) numbers, respectively. The LTI dynamical system $\mathbf K$ pencil is denoted $\Lambda_{\mathbf{K}}$. We denote with $h\in\Real_+$, the sampling-time, with $I$ the identity matrix and $\imath$ ($\sqrt{\imath}=-1$) the complex variable. The time average of a quantity is denoted $\langle \rangle$. Regarding the model, $c_{\text{flap}}$ and $c_{\text{tot}}$ respectively stand for the flap chord length and the model total length. The flap deflection angle is denoted $\delta$. The freestream velocity is noted $U_\infty$ and the Reynolds number based on the model total length is computed as $Re=U_\infty c_{\text{tot}}/\nu$, with $\nu$ the air kinematic viscosity. Considering the actuators, $f$, $f^+$ and $\alpha$ refer to the actuation frequency, its reduced form and to the duty cycle. The momentum coefficient $C_\mu$ is derived from $q_{\text{jet}}$, $U_{\text{jet}}$, $\rho_\infty$ and $A_{\text{ref}}$ respectively the actuators mass flow rate, actuators outlet velocity, freestream density and reference area chosen as the flap area. Concerning the sensors setup, the pressure coefficient $C_p$ calculation is based on the static pressure $p$ and the freestream pressure $p_\infty$. The flap lower and upper surfaces pressure coefficients are denoted $C_{p_{\text{lower}}}$ and $C_{p_{\text{upper}}}$. These coefficients help computing the lift coefficient $C_L$ as detailed in the following. In some of the following figures considering the controlled case, zones without and with actuation are distinguished. They respectively denote cases in which the control is applied but valves are not actuated or actuated with a duty cycle fixed by the controller.


\section{Experimental control setup description}
\label{sec:setup}
\subsection{Setup overview and wing surface properties}

Flow control experiments were carried out in the L1 wind tunnel facility at ONERA, Lille. It is characterized by a test section diameter of 2.40 m. The model is composed of a 867 mm long flat plate, stabilizing the boundary layer, followed by a plain flap of chord length $c_{\text{flap}}=220$ mm, yielding a model total length $c_{\text{tot}}=1087$ mm. The plain flap design is based on a NACA 4412 airfoil. As depicted in Figure \ref{fig:setup}, two lateral panels separated by 800 mm are placed inside the wind tunnel test section, next to the model borders to avoid side effects and ensure that the flow developing over the model is bi-dimensional. The flow separation control over this model has already been studied by Chabert \textit{et al} \cite{chabert_experimental_2014,chabert_experimental_2014-1}, who implemented a slope seeking control algorithm. The boundary layer developing over the model is turbulent. Its transition is triggered by a carborundum line placed at the flat plate leading edge. Velocity measurements upstream the model are performed with Pitot tubes. During the experiments, $U_\infty$ is fixed to 34.5 m/s, yielding $Re=2.39 \times 10^6$. The turbulence level inside the wind tunnel is about $1.3\ \%$. The model is placed inside the test section with a $0^\circ$ angle of attack. The motorized flap can be deflected downward at an angle $\delta$ varying between $2^\circ$ and $37^\circ$. Note that the flap is not an actuator used in the control loop but a way to modify the configuration accounting for uncertainties and allowing exhibiting the separation phenomena (Figure \ref{fig:performanceObjective}).
%


\subsection{Actuators and sensors description}

\subsubsection{Actuation system}

The actuation setup is composed of 7 slots spanning along the flap leading edge at a location of $x/c_{\text{flap}}=0.08$. Separated by 7 mm from each other, the actuators slots are 90 mm long and 0.25 mm thick each. They cover 80\% of the flap span and are supplied by on/off Festo MHE2 fast response solenoid valves, fed with pressurized air up to 7 bar. The slot outlet velocity is inclined by $30^\circ$ relatively to the flap local tangent. The valves mass flow rate and the actuation frequency are fixed respectively to $21$ g/s and to $f=100$ Hz. The reduced actuation frequency referred as $f^+$ is therefore:
\begin{equation}
    f^+ = \dfrac{f\ c_{\text{flap}}}{U_\infty} \cong 0.64
\end{equation}
The added momentum to the flow can be characterized by $C_\mu$ defined as:
\begin{equation}
    C_\mu = \dfrac{q_{\text{jet}}U_{\text{jet}}}{\dfrac{1}{2}\rho_\infty U_\infty^2 A_{\text{ref}}},
\end{equation}
Considering the conditions detailed above, the constant blowing momentum coefficient has a value of 1.6\%.

As detailed in \cite{chabert_these}, for a fixed mass flow rate supplied to the actuators, the momentum coefficient can be extended to pulsed blowing according to the following equation:
\begin{equation}
    {\langle C_\mu \rangle = \dfrac{1}{\alpha} \dfrac{\rho_{\text{jet}} \langle U_{\text{jet}}^2 \rangle A_{\text{jet}}}{\dfrac{1}{2}\rho_\infty U_\infty^2A_{\text{ref}}}},
\end{equation}
As the valves considered here are driven with square signals with a duty cycle $\alpha$, the momentum coefficient can be simplified as:
\begin{equation}
    {\langle C_\mu \rangle = \dfrac{1}{\alpha} \dfrac{q_{jet}U_{jet}}{\dfrac{1}{2}\rho_\infty U_\infty^2 A_{ref}} = \dfrac{1}{\alpha}C_\mu}.
\end{equation}

The $\alpha$ variable then denotes the control signal used in the loop control. During the flow control experiments $\langle C_\mu \rangle$ used reached a maximum value of 4.6\%. 

\subsubsection{Sensing system} 
Regarding the sensors setup, 51 pressure taps are dispatched on both the flat plate and flap upper and lower surfaces. From these pressure measurements both the pressure coefficient at each tap location and the global lift coefficient can be computed. The pressure coefficient is defined according to the following equation:
\begin{equation}
    {C_p = \dfrac{p-p_\infty}{\dfrac{1}{2}\rho_\infty U_\infty^2}},
\end{equation}
The lift coefficient is derived from the pressure coefficient computations according to the following formula:

\begin{equation}
    {C_L = \int_{0}^{1} (C_{p_{\text{lower}}}-C_{p_{\text{upper}}}) \, {d}\dfrac{x}{c_{\text{tot}}}}, 
\end{equation}
Both quantities $C_p$ and $C_L$ are used to assess the control effects, comparing cases of uncontrolled and controlled flows.

\subsubsection{Experimental strategy} 

\begin{figure}[h]
    \centering
    \includegraphics[width=.8\columnwidth]{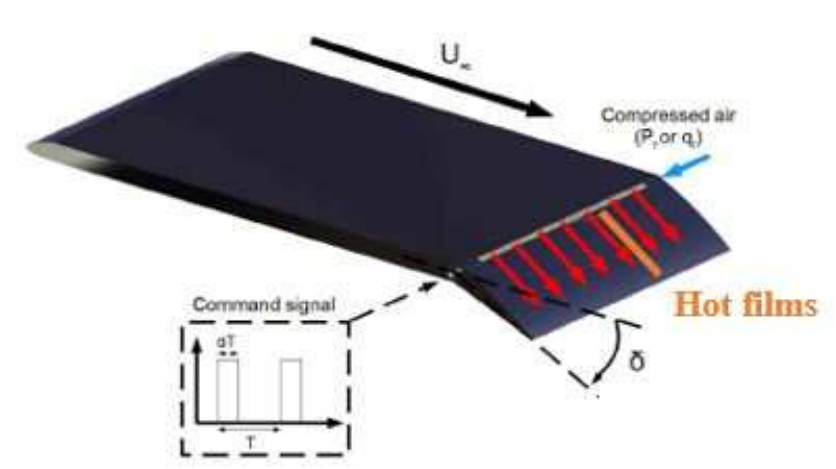}
    \caption{Scheme of the model placed in the wind tunnel with the actuators command and hot films positions.}
    \label{fig:actuators_sensors}
\end{figure}

To monitor the flow separation over the flap and implement the control part, eight Senflex\textsuperscript{\textregistered} hot films are placed along the flap chord-wise direction. Connected to two Dantec\textsuperscript{\textregistered} Streamlines units, hot films signals are recorded at a sampling frequency of $1.25$ kHz over 3 minutes for each measurement points. The control tracking value is defined with respect to the fifth hot film measurements, located at the dimensionless abscissa $x/c_{\text{flap}}=0.511$, taking the coordinate origin at the flap leading edge. Control scripts, written in LabVIEW Real-Time 2011 via a PXIe-8102 controller, are embedded in a PXI chassis. Both the actuators and sensors setups are sketched in Figure \ref{fig:actuators_sensors}.

%


\subsection{Performance characterisation toward specifications}

\begin{figure}[h]
    \centering
    \includegraphics[width=\columnwidth]{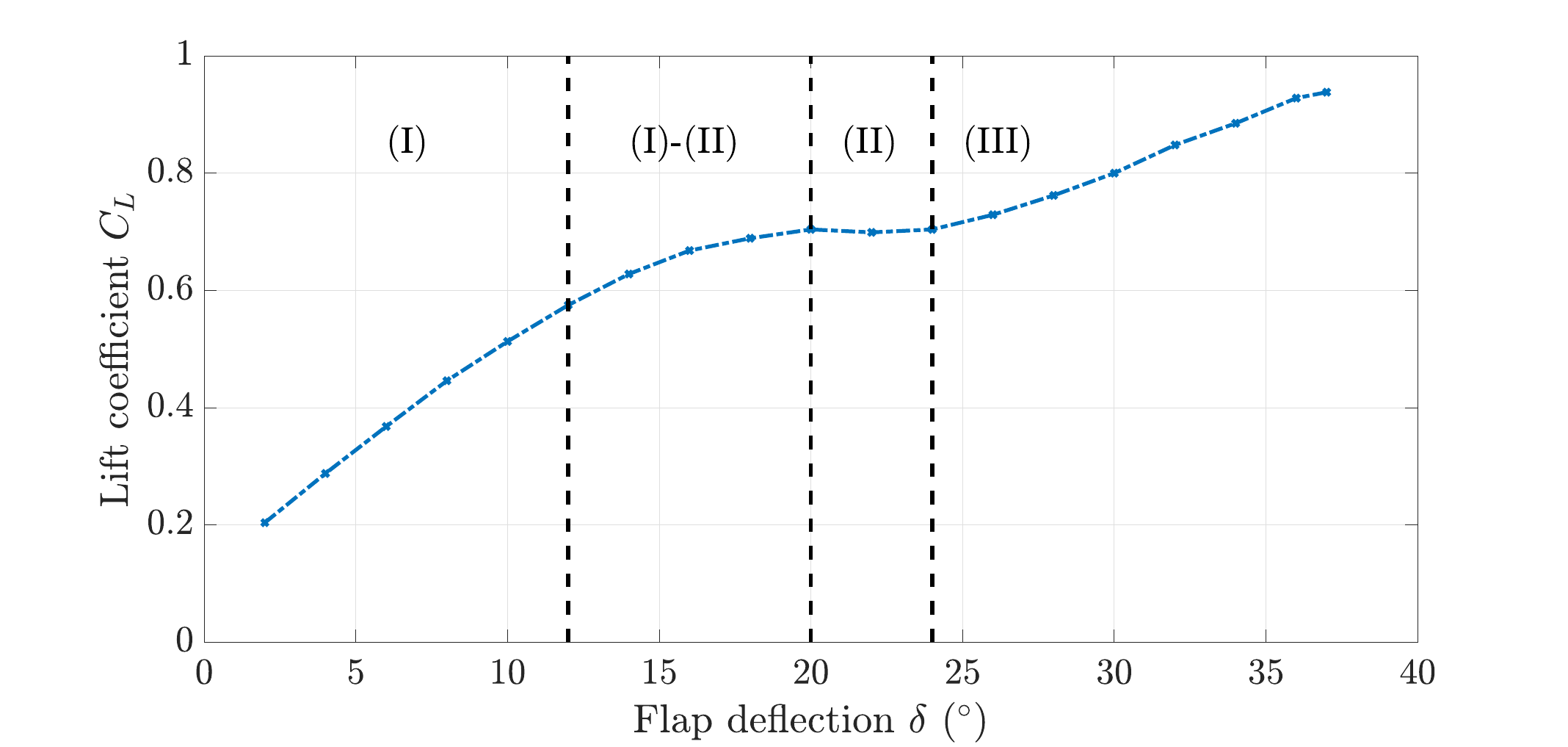}
    \caption{Evolution of the unforced flow lift coefficient $C_L$ against the flap deflection angle $\delta$ ($U_\infty = 34.5$ m/s).}
    \label{fig:lift_unforced}
\end{figure}

First measurements focused on the unforced flow characterization. The freestream velocity was fixed to $U_\infty =34.5$ m/s and the flap was deflected from $2^\circ$ down to $37^\circ$. The evolution of the lift coefficient is presented in Figure \ref{fig:lift_unforced} and similar results to the one presented in \cite{chabert_experimental_2014-1} are obtained. 

Following Figure \ref{fig:lift_unforced} and as described by Hoerner \cite{hoerner_fluid-dynamic_1985}, four zones can be distinguished. The first zone (I), for $\delta$ between $2^\circ$ and $12^\circ$, describes a linear evolution of $C_L$ against $\delta$. The second zone (I)-(II) corresponds to a slower increase in the lift coefficient and spreads between $12^\circ$ and $20^\circ$, indicating the development of the flow separation over the flap. The zone denoted (II) corresponds to a plateau of $C_L$ due to the recirculation bubble entirely developed over the flap. Finally, the zone (III) denotes a zone of non-linear increase in the lift coefficient. The non-linear evolution of $C_L$ in (III) would be better observed with higher deflection angles, as pointed out by Hoerner \cite{hoerner_fluid-dynamic_1985}.

\begin{figure}[h]
    \centering
    \includegraphics[width=\columnwidth]{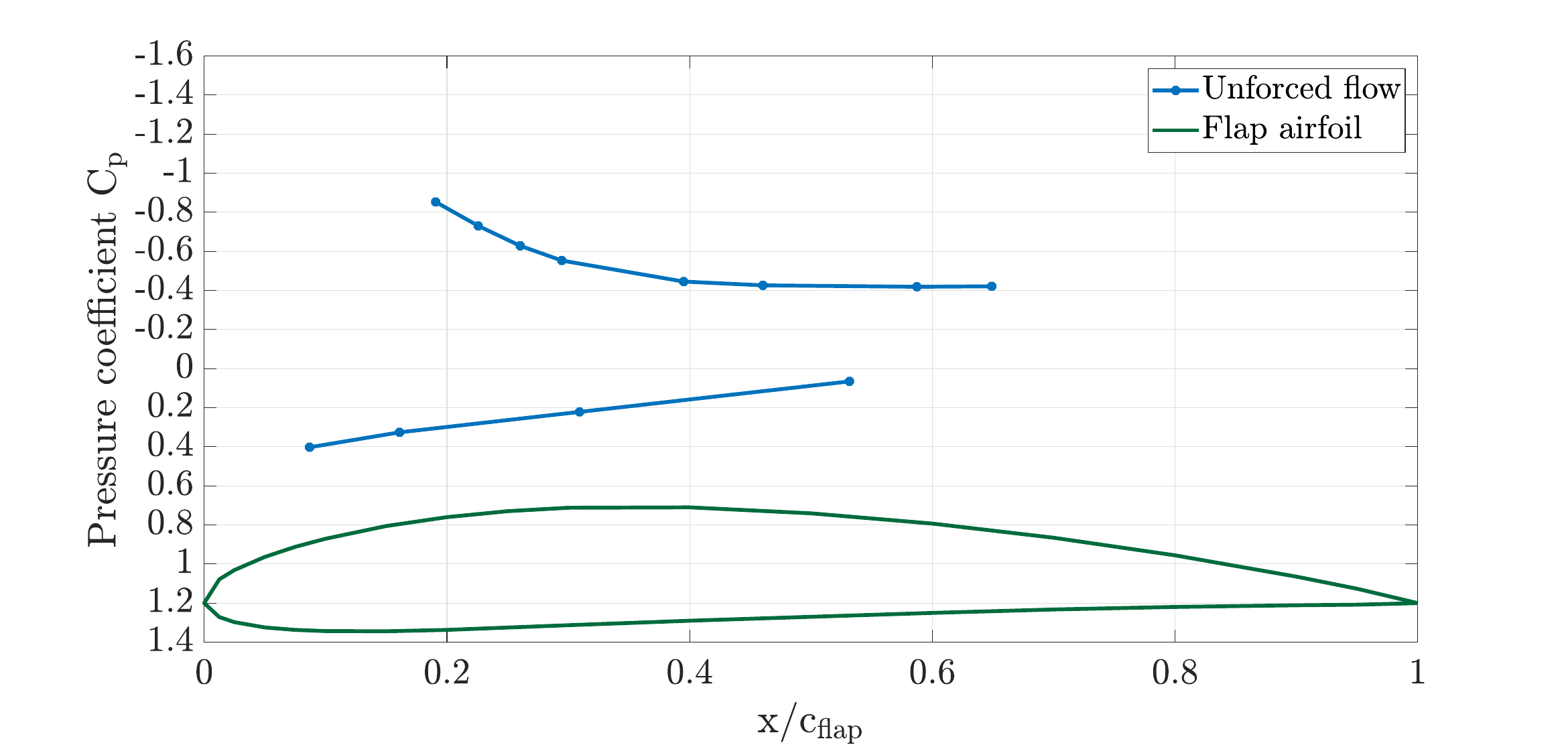}
    \caption{Evolution of the unforced flow pressure coefficient $C_p$ against the dimensionless abscissa $x/c_{\text{flap}}$. ($\delta =18^\circ$ and $U_\infty = 34.5$ m/s). The two curves represent the profile upper and lower coefficients.}
    \label{fig:Cp_18deg}
\end{figure}

The development of the recirculation area over the flap can also be observed in the pressure coefficient. Figure \ref{fig:Cp_18deg} highlights the evolution of $C_p$ against the dimensionless abscissa $x/c_{\text{flap}}$ for a deflection angle of $18^\circ$. The pressure gradient between 0.19 and 0.46 is followed by a plateau of $C_p$ between 0.46 and 0.65, indicating the flow separation. The longer the plateau is, the longer the flow recirculation area is. Therefore, as the flap deflection angle is increased, the $C_p$ plateau spreads over a larger area.

\begin{figure}[h]
    \centering
    \includegraphics[width=\columnwidth]{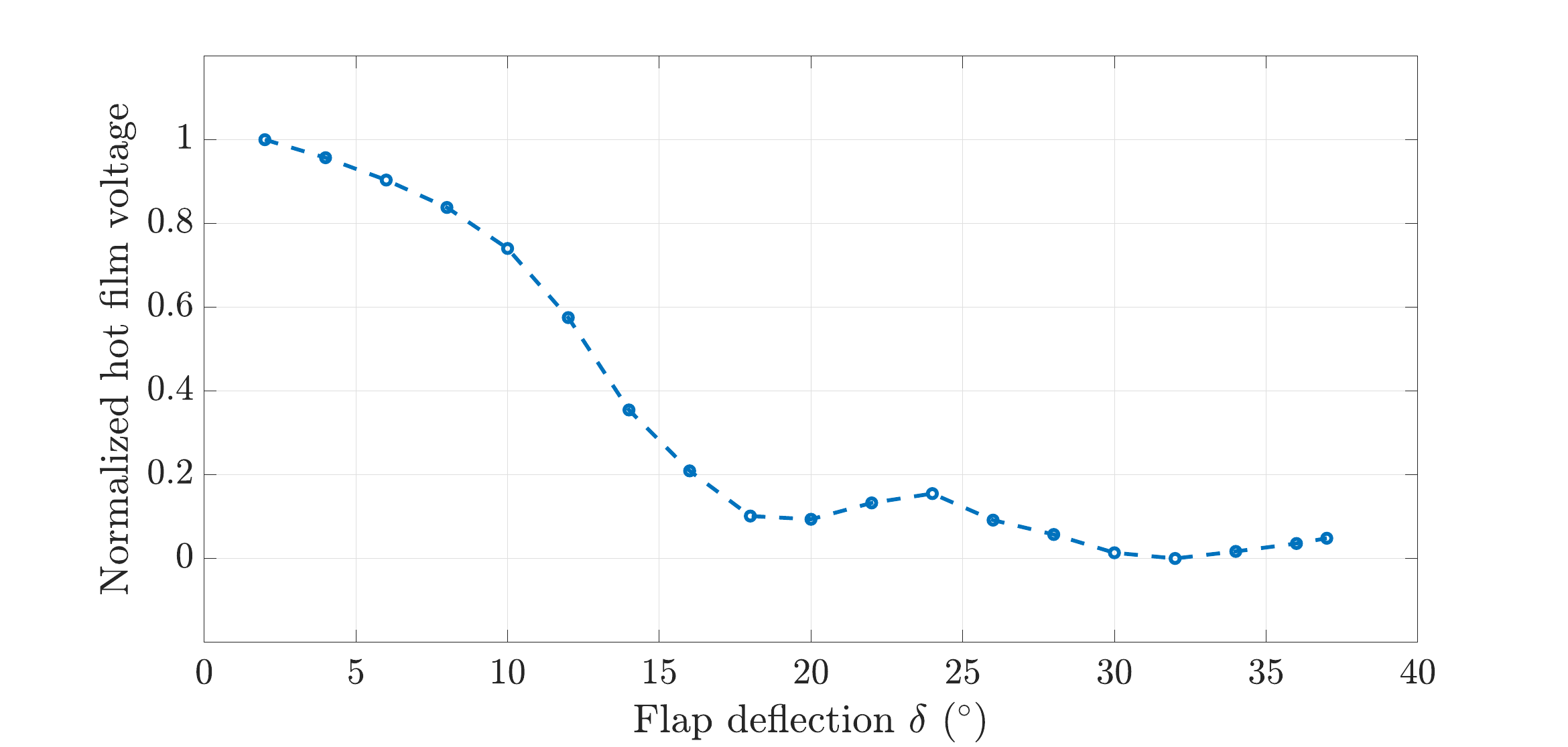}
    \caption{Evolution of the normalized voltage of the fifth hot film against the deflection angle $\delta$ ($U_\infty = 34.5$ m/s).}
    \label{fig:HF_5_normalized}
\end{figure}

Signals of the 8 hot film sensors were also recorded during these tests. For each deflection angle, hot films time series have been averaged and normalized according to $
U^\ast = (\langle U \rangle -U_{min})/(U_{max}-U_{min})$, where $U^\ast$ is the dimensionless hot film voltage, $\langle U \rangle$ the hot film mean voltage, $U_{min}$ and $U_{max}$ respectively are the hot film minimum and maximum voltages. Figure \ref{fig:HF_5_normalized} depicts the evolution of $U^\ast$ for the fifth hot film on the flap, which is located at $x/c_{\text{flap}}=0.511$. As the flap angle is increased the hot film normalized voltage decreases. This trend indicates a decreasing wall shear stress, which reaches a local minimum for $\delta = 20^\circ$. The first minimum at this flap angle, points out the flow separation at this location. When the flap is further deflected, the hot film voltage increases as the flow recirculation bubble intensifies and reaches a local maximum. The voltage reaches a global minimum for an angle of $\delta =~ 32^\circ$. This second minimum may be due to the apparition of a second recirculation zone occurring for high deflection angles, as observed in \cite{chabert_these}.

\subsection{Reference signal and control architecture}

\begin{figure}[h]
    \centering
    \includegraphics[width=\columnwidth]{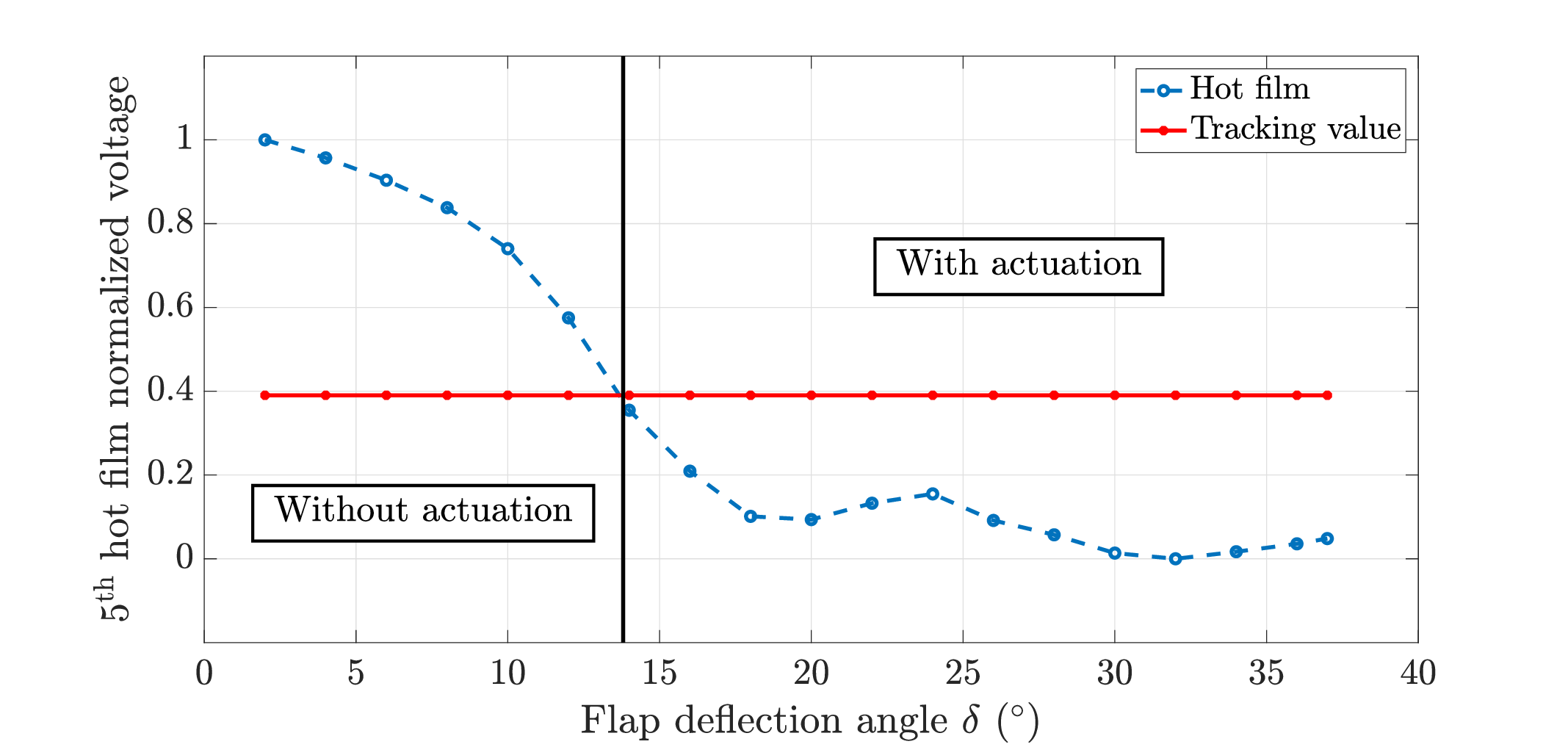}
    \caption{Evolution of the normalized voltage of the fifth hot film (blue curve) and the reference value (red curve) against the deflection angle $\delta$ ($U_\infty = 34.5$ m/s). The black curve separates areas without and with actuation.}
    \label{fig:Control_definition_HF5}
\end{figure}

Following these observations, a reference value for the fifth hot film sensor is defined, such that the flow separation over the flap is avoided. The normalized objective value $U_{\text{obj}}^\ast$ is fixed to $0.3903$. In Figure \ref{fig:Control_definition_HF5}, this reference value is superimposed with the evolution of the fifth hot film normalized voltage. The controller aim is therefore to maintain the hot film normalized voltage to the reference value. Therefore, two different zones are defined in this chart. One for deflection angles $\delta < 13.8^\circ$ and the second one for $\delta > 13.8^\circ$. The first one corresponds to deflection angles for which actuators do not add momentum to the flow, as the hot film voltage is above the reference value. The second one defines angles $\delta$ for which the Festo valves are cyclically actuated and aim at maintaining the hot film voltage to the reference value. Based on this reference value, from now on denoted $\mathbf r$, ability of both control strategies (either linear data-driven or positive model-driven) to maintain this value is investigated.

\begin{figure*}
    \centering
    \scalebox{.6}{\input{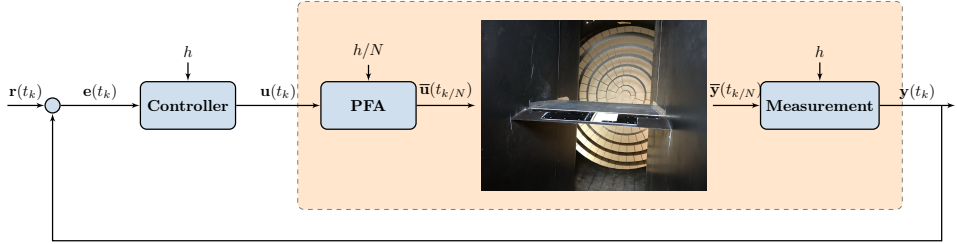}}
\caption{Overview of the considered closed-loop architecture. The \textbf{controller}, sampled at frequency $h$, feeds a series of PFA acting along the wing span. The system is illustrated by the setup photo, and the measurement is achieved by the hot films located along the wing flap. The orange block is the overall system.}
    \label{fig:closedLoopScheme}
\end{figure*}

Based on the above considerations, a \emph{reference signal tracking feedback control architecture} can be set up. With reference to Figure \ref{fig:closedLoopScheme}, one aims at designing a $h$-sampled control law aiming at ensuring that the output signal $\mathbf y(t_k)$, measured by the fifth hot film, tracks $\mathbf r(t_k)$, the reference level previously defined. The controller provides a sampled-time continuous order $\mathbf u(t_k)$, transformed in an on-off one $\overline{\mathbf u}(t_{k/N})$, leading to the controlled duty cycle $\alpha$ applied by the PFA (Pulsed Fluidic Actuator), sampled $N$ times faster (see also \cite{PoussotCST:2022} for details on the PFA). The control problem boils down to a \emph{reference tracking one}.

\begin{remark}[Sensor location]
In the considered experimental case, the fifth sensor is selected. However, other locations or even multiple sensors may be considered. The impact of this placement/selection may be considered in future works. The choice for this sensor was dictated by the following considerations: first, the quality of the signal was good, second, it was located far enough to actually see the separation phenomenon.
\end{remark}

\begin{remark}[Control architecture extensions]
Similarly, in the considered experimental context, a single input, single output controller is sought. The rest of the section sticks to this configuration. However, extensions to multi-input and single-output are possible. Extensions to multiple actuators are also possible but would lead to considerably more complex analysis. 
\end{remark}

\section{Flow separation control tuning}
\label{sec:control}
Considering the control architecture in Figure \ref{fig:closedLoopScheme}, this section details the design and tuning of the controller. Two different strategies are implemented and evaluated. First, the linear Loewner Data-Driven Control (L-DDC) \cite{KergusPhD:2019} (section \ref{LDDC_design}) and second, a phenomenological-driven nonlinear positive control \cite{BriatSIADS:2020} (section \ref{Positive_design}). Both configurations performances are commented and illustrated in section \ref{ssec-flow_result}. 

\subsection{L-DDC design}
\label{LDDC_design}

\subsubsection{Idea and principle}

The L-DDC belongs to the so-called data-driven reference model approaches\footnote{DDC methods have a long history dating to the proportional, integral, derivative (PID) tuning method by Ziegler-Nichols in early 40's or the self tuning regulator by \AA str\"om in the 90's (see \eg \cite{Ziegler:1942} for more details and references). This field remains still very active (see \eg \cite{Formentin:2014})}. 
The {L-DDC} procedure boils down to two steps: first deriving the \emph{ideal controller} denoted $\mathbf K^\star$, and second, the \emph{controller identification} via interpolation in the Loewner framework \cite{Mayo:2007,GoseaHNA:2022}. 

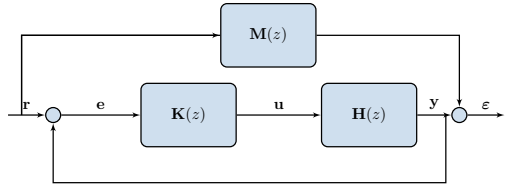
\begin{figure}[h!]
\centering
\scalebox{.6}{\tikzstyle{block} = [draw, thick,fill=bleuONERA!20, rectangle, minimum height=4em, minimum width=6em,rounded corners]
\tikzstyle{sum} = [draw, thick,fill=bleuONERA!20, circle, node distance=1cm]
\tikzstyle{input} = [coordinate]
\tikzstyle{output} = [coordinate]
\tikzstyle{pinstyle} = [pin edge={to-,thick,black}]
\tikzstyle{connector} = [->,thick]

\begin{tikzpicture}[auto, node distance=3cm,>=latex']
    \node [input, name=input] {};
    \node [sum, right of=input] (sum) {};
    \node [block, right of=sum] (controller) {$\mathbf K(z)$};
    \node [block, right of=controller, node distance=4cm] (system) {$\mathbf H(z)$};
    \node [sum, right of=system, node distance=2cm] (sum2) {};
    \node [block, above right of=controller, node distance=2.5cm and 2cm] (obj) {$\mathbf M(z)$};
    \draw [connector] (controller) -- node[name=u] {${\bu}$} (system);
    \node [output, right of=sum2, node distance=1cm] (output) {};

    \draw [connector] (input) -- node {${\mathbf r}$} (sum);
    \draw [connector] (sum) -- node {$\mathbf e$}(controller);
    \draw [connector] (system) -- node [name=y] {${\mathbf y}$}(sum2);
    \draw [connector] (sum2)+(-0.3cm,0) -- ++(-0.3cm,-1.5cm) -| node [near start] {} (sum.south);
    \draw [connector] (sum2) -- node {$\boldsymbol \varepsilon$}(output);
    \draw [connector] (input)+(0.3cm,0) |- (obj) -| (sum2);
    \draw [connector] (input)+(0.3cm,0) |- (obj);
    
\end{tikzpicture}}
\caption{Data-driven control problem formulation. $z$ denotes the complex variable either in the continuous or sampled-time.}
\label{fig:ddc_problem}
\end{figure}

We recall the main steps in the SISO case and with the considered reference tracking architecture. Following Figure \ref{fig:ddc_problem}, the objective is to find an LTI controller with transfer function $\mathbf{K} : \Cplx\backslash\Lambda_{\mathbf{K} }\rightarrow\Cplx$ that minimizes the transfer difference between $\mathbf r$ and $\boldsymbol{\varepsilon}$, \ie between the resulting closed-loop and a user-defined reference model $\mathbf{M}: \Cplx\backslash\boldsymbol\Lambda_{\mathbf{M}}\rightarrow\Cplx$. This is made possible through the definition of the ideal controller $\mathbf{K}^\star$, being the {LTI} controller that would have given the desired reference model behaviour if inserted in the closed-loop. The latter is defined as $\mathbf{K}^\star=\Htran^{-1}\mathbf{M}(I-\mathbf{M})^{-1}$, where $\Htran:\Cplx\backslash\boldsymbol\Lambda_{\Htran}\rightarrow\Cplx$ is the model of the system to control. In the data-driven case, when $\Htran(z)$ is not explicitly known but may be evaluated at some frozen values $z_k\in\Cplx$, this definition may be recast as a set of $k=1,\dots,N$ equations:
\begin{equation}
    \mathbf{K}^\star(z_k)=\boldsymbol\Phi_k^{-1}\mathbf{M}(z_k)(I-\mathbf{M}(z_k))^{-1},
    \label{eq:Kideal}
\end{equation}
where $\boldsymbol\Phi_k=\Htran(z_k)\in\Cplx$ is the evaluation of the unknown model at $z_k$. In an experimental context, one usually considers sampling $\Htran$ at $z_k=\imath \omega_k$ ($\omega_k\in\Real_+$). In this case, $\boldsymbol\Phi_k$ is the frequency response of the open-loop system at $\omega_k$. Then, the couple 
\begin{equation}
    \{z_k,\mathbf{K}^\star(z_k)\}_{k=1}^N,
    \label{eq:raw_data}
\end{equation}
is referred to as the \emph{raw data} for our controller design. Finding a controller $\mathbf K$ that \emph{fits} \eqref{eq:raw_data} can be considered to be an identification/interpolation problem which may be solved by many approaches. The Loewner framework \cite{Mayo:2007} allows constructing both a function $\mathbf K$ with minimal McMillan degree and realization order $n\leq N$, satisfying conditions \eqref{eq:Kideal} or an approximation of it with a realization of order $r<n$. 

\begin{remark}[Advantages of the L-DDC]
L-DDC is a combination of determining the ideal controller from frequency-domain data via a reference model and the use of the Loewner framework to construct a reduced order controller. Such an interpolatory-based data-driven control design solves problems faced by practitioners: {(i)} the controller design is directly obtained using open-loop raw data \eqref{eq:raw_data} collected on the experimental setup, {(ii)} without any optimization process, only linear algebra manipulation, {(iii)} and without any prior controller structure or order specification (these latter may be automated by a rank revealing factorization). This approach has proven to be effective for digital control \cite{VuilleminWC:2020} and on experimental application \cite{PoussotCST:2022}. \cite[sec. 4]{GoseaHNA:2022} provides practical details and didactic examples including infinite dimensional systems. 
\end{remark}

\begin{remark}[Limitations of the L-DDC]
As a linear controller design, it embeds regular linear limitations. One important has been highlighted during the experiments. It concerns the fact that it does not handle actuator limitations, which is in this case works on/off only and are only able to blow air. In presence of an integral action, this may result in stability issues. This point is discussed in sections \ref{Positive_design} and \ref{ssec-flow_result}.
\end{remark}

\subsubsection{Application} 

\begin{figure}[h]
    \centering
    \includegraphics[width=.8\columnwidth]{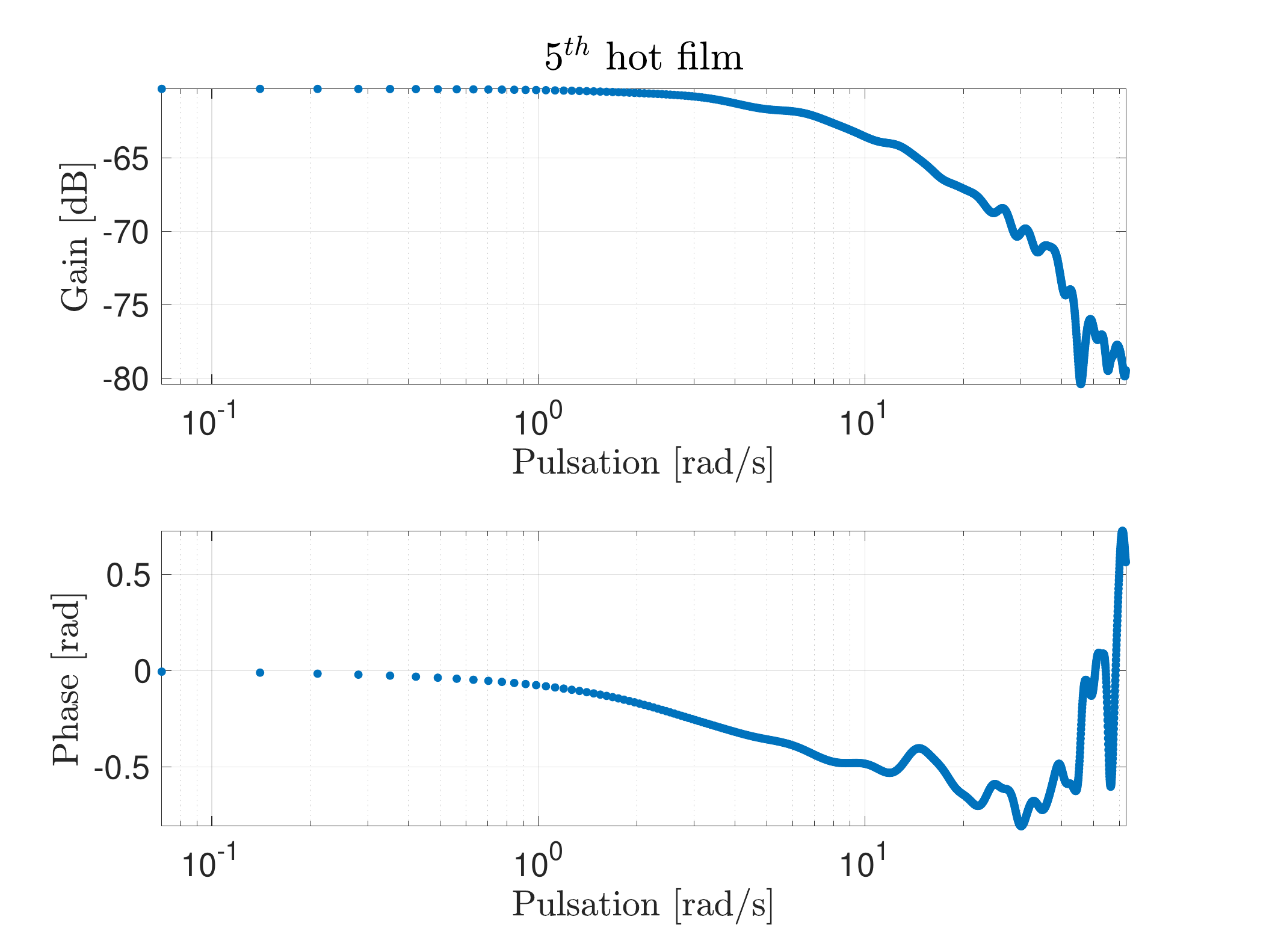}
    \caption{Frequency response gain and phase diagrams of the data $\boldsymbol \Phi_k$ collected during the open-loop experiments.}
    \label{fig:Frequency_data}
\end{figure}

The frequency-domain response describing the separated flow dynamics over the flap is first needed. The operating point for this step is a deflection angle $\delta=24^\circ$ and $U_\infty=34.5$ m/s, operating point for which the flow is indeed separated at the considered measurement position. The actuators command signal $\bu(t_k)$ consists in a logarithmic frequency sweep applied to the duty cycle with frequencies ranging from $0.01$ Hz to $10$ Hz over $180$ s.
From this input, the fifth hot film response $\by(t_k)$ is collected. The discrete frequency-domain transfer data from $\bu(t_k)$ to $\by(t_k)$ is obtained and denoted $\{\imath\omega_k,\boldsymbol\Phi_k\}_{k=1}^N$, where $\omega_k\in\Real_+$ is the pulsation and $\boldsymbol\Phi_k\in\Cplx$ is the SISO transfer response of the system (orange block in Figure \ref{fig:closedLoopScheme}) and $N\in\mathbb N$ is the length of the FFT. The data-driven Bode-like diagram is presented in Figure \ref{fig:Frequency_data}. It exhibits a gain drop around 1 rad/s and a decay in the phase, characteristic of delayed and fractional systems\footnote{Note that from this point, model identification may be done in order to apply model-driven method. Here we skip this step to directly go to the control design.}.

\begin{figure}[h]
    \centering
    \includegraphics[width=.8\columnwidth]{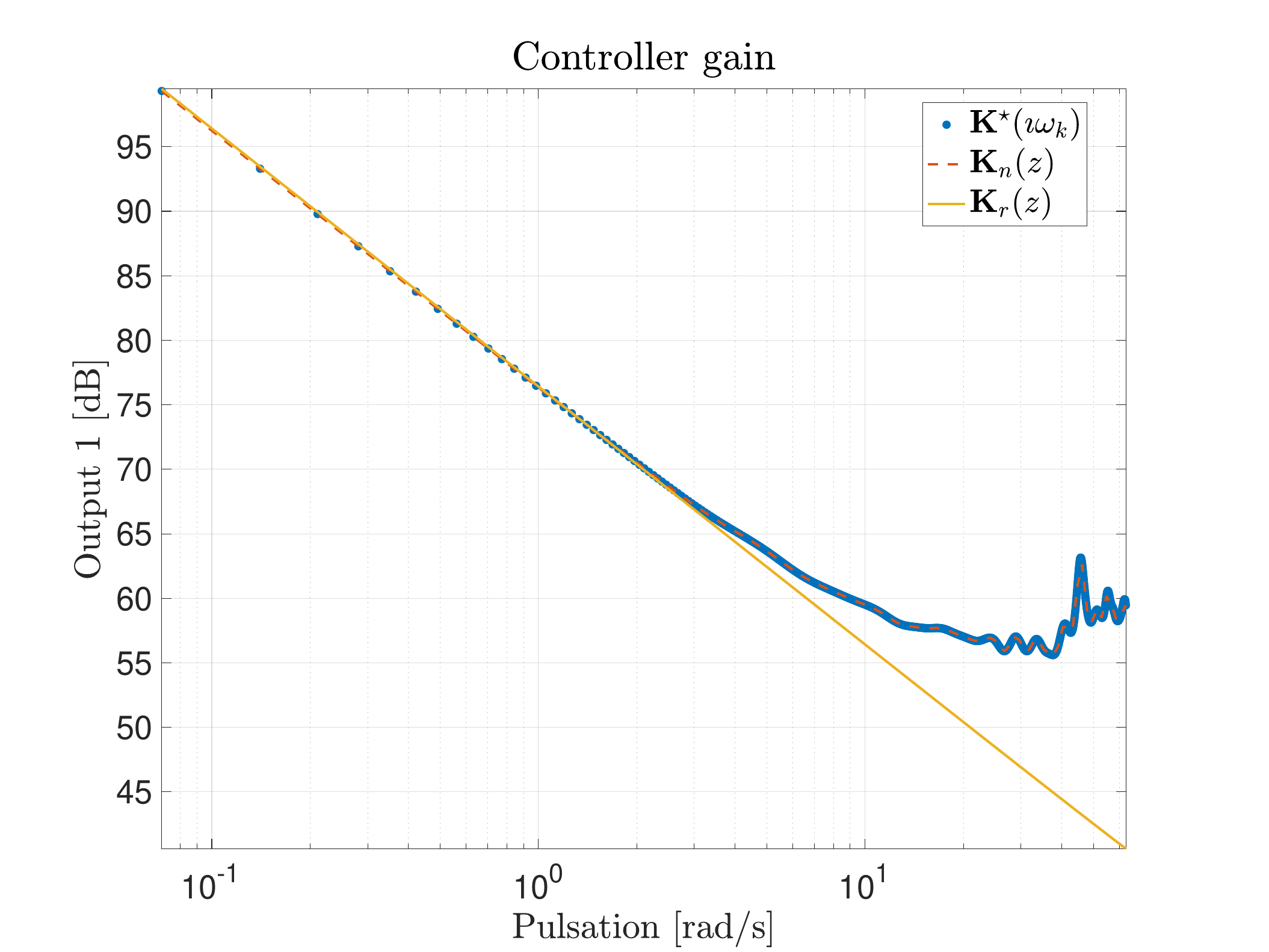}
    \caption{Bode gain diagrams of the ideal controller data $\mathbf K^\star$ \eqref{eq:Kideal}, its exact interpolated sampled-time controller $\mathbf K_n$ and its approximation $\mathbf K_r$ with an order $r=1$.}
    \label{fig:ddc_controller}
\end{figure}

Simultaneously with the previous step, the objective closed-loop transfer function $\mathbf M$ is defined as a first order model $\mathbf M(s)=1/(s/w_0+1))$ where $\omega_0=2\pi$ rad/s, is the natural cut-off frequency. We refer to the black dashed lines of Figure \ref{fig:performanceObjectiveFR}, given in the introduction. $\mathbf{M}$ mainly aims at ensuring no steady-sate error (static gain objective set to one)\footnote{Note that the selection of a first order reference model may be subject to discussions. Here we aim at illustrating the most simple settings. For more details on the choice of $\mathbf M$, reader is invited to refer to \cite{KergusPhD:2019}.}. With reference to equation \eqref{eq:Kideal} we are now ready to compute the ideal controller $\mathbf K^\star$ as well as its exact interpolation $\mathbf K_n$, where $n=128$ is automatically selected by the rank revealing factorization embedded in the Loewner process, and its approximations $\mathbf K_r$ with an order $r=1$. After time-domain discretisation ($h=1/100$ s), Figure \ref{fig:ddc_controller} illustrates the controller frequency response gains. The implemented linear controller is a pure sampled-time integrator with gain $k=66.19$ (\ie $\mathbf K_r(z)=66.19/(z-1)$). Obviously, a proportional integral action model may also be identified with a better accuracy. Here we stick to the integral action in view of the nonlinear integral control analysis analysed in the next section.

Figure \ref{fig:ddc_controller} well illustrates that the exact interpolation perfectly matches the data and the approximation with an order $r=1$ preserves the integral term. Such an observation, coupled with the knowledge of the system input-output positivity property (input blows air and output measures positive values only), motivates the use of a more involved control strategy discussed in the next section.

\subsection{Nonlinear positive control design}
\label{Positive_design}

\subsubsection{Idea and principle} 

Observing that the considered system is stable (indeed, the configuration is an amplification one, but without any instability) and input-output positive (\ie for any nonnegative input $\mathbf u$, the output $\mathbf y$ is nonnegative), it seems interesting to exploit this property for control purposes. Although not recent \cite{Farina:00}, positive systems have recently attracted a lot of attention due to their surprising properties; see e.g. \cite{Briat:11h,Ebihara:11}. In particular, the theory of (linear) positive systems is playing an essential role in the modeling, the analysis and the control of compartmental systems which include biological, physiological, epidemiological and ecological systems as special cases \cite{Haddad:10}. Recently, a novel type of integral controller -- the Antithetic Integral Controller (AIC) -- was introduced in the context of biological control and chemical reaction networks \cite{BriatCellSystems:2016}. The rationale for introducing such a controller was, among others, the derivation of a controller having a positive system representation that could always return a nonnegative control input. It was later proved in \cite{BriatSIADS:2020} that this nonlinear integral controller enjoyed certain interesting properties which are absent from its linear (\ie non-positive) counterpart. It is also worth mentioning that other nonlinear positive integral controllers exist \cite{BriatSIADS:2020}, but the AIC exhibits a lot of the desirable behavioral properties of the usual integral controllers and this is the reason why it is considered here. Indeed, \cite[Thm. 3.6]{BriatSIADS:2020} provides a stability proof of the closed-loop interconnection if the original underlying model is a linear positive one. In fact, those stability conditions coincide, in the worst-case, with the stability conditions of the standard integral controller, which indicates that using the AIC is not more constraining than using a linear integral controller. 

\begin{remark}[Closed-loop stability]
In the L-DDC setting, no stability proof can be guaranteed a-priori. This may be checked afterward with specific data-driven techniques (see \eg \cite[chap. 7]{KergusPhD:2019} or \cite{PoussotGTA:2022}). However, one important feature of the positive design by Briat \cite[equation 2.1]{BriatSIADS:2020} is that the closed-loop control of a stable positive system with an AIC ensures local exponential stability, while respecting input signal constraints, under very mild conditions. 
\end{remark}

\subsubsection{Application} 

The original AIC is given in \cite[equation 2.1]{BriatSIADS:2020} with the following equation set, 
\begin{eq}
\begin{array}{rcl}
\dot z_1(t) &=& \mathbf r(t) - \eta z_1(t) z_2({t}) \\
\dot z_2(t) &=& \by({t}) - \eta z_1(t)z_2(t) \\
\bu({t_k})   &=& k z_1(t)
\end{array}.
\label{eq:positive_th}
\end{eq}
The discretized version (using the backward method)  takes the form ($h=1/100$ s):
\begin{eq}
\begin{array}{rcl}
z_1(t_k+h) &=& z_1(t_k) + h\big(\mathbf r({t_k}) - \eta z_1({t_k}) z_2({t_k})\big) \\
z_2(t_k+h) &=& z_2(t_k) + h\big(\by({t_k}) - \eta z_1(t_k)z_2(t_k)\big) \\
\bu({t_k})   &=& k z_1(t_k)
\end{array}.
\label{eq:positive}
\end{eq}

Implemented in the real-time environment, user then tunes the values of $k\in\Real$ and $\eta\in\Real_+$ according to the desired controller. As \eqref{eq:positive} aims at reproducing the integral action, in our setting, gain $k$ has been set equal to the integral term obtained with the L-DDC approach; \ie $k=66.19$  (section \ref{LDDC_design}) and $\eta=300$, used to tend to a pure integral action. We refer to \cite{BriatSIADS:2020} for further details. 

\subsection{Flow control experimental results}
\label{ssec-flow_result}

To validate the proposed reference feedback (linear and nonlinear) integral control, four different type of experiments are carried out. In \S\ref{ssec:lift}, the lift coefficient $C_L$ with feedback is computed and compare to the uncontrolled one (notice that this is the utilate objective of the control). In \S\ref{ssec:delta}, the robustness of the control with the deflection angle is analysed. In \S\ref{ssec:freq}, the frequency response of the controlled flap is computed and compared to the expected objective. Finally in \S\ref{ssec:positive}, some considerations on the nonlinear positive controller are discussed.

\subsubsection{Lift coefficient gain evaluation ($C_L$)}
\label{ssec:lift}

Both linear and non-linear integral controllers have been applied for the same flow conditions. As expected for deflection angles below $13.8^\circ$, valves are not opened as the hot film voltage is above the reference value $\mathbf r$. When the hot film voltage tends to stand below the reference value, valves are opened with a duty cycle determined by the controller (this is typically the case when flap angle is increased). As an effect, the hot film voltage is maintained at the reference value thanks to the feedback control action. As presented in the introduction, Figure \ref{fig:performanceObjective} highlights the benefit of control on the lift coefficient $C_L$ increase. 

For $\delta < 13.8^\circ$, both uncontrolled and controlled flow present the same lift coefficient, as in both cases valves are not opened. However, for $\delta > 13.8^\circ$, curves for the uncontrolled and controlled cases do not superimpose anymore. The lift coefficient in the controlled case is higher than the one of the uncontrolled case. Regarding the controlled case, the linear evolution of $C_L$ is extended up to $\delta =24^\circ$. Between $\delta = 24^\circ$ and $\delta = 26^\circ$, $C_L$ is reduced drastically. This phenomenon is due to the apparition of flow separation at the flap trailing edge. This flow separation does not spread over the entire flap as the control counters it and hold it at the flap trailing edge\charles{\footnote{This observation motivates future investigations involving additional sensors and a more involved multi-input control.}}. These results can also be observed in the analysis of the pressure coefficient evolution in both the uncontrolled and controlled flow cases. As both the linear controller and the non-linear positive controller have been applied with the same reference value on the hot film, both control cases yielded the same results on the lift and pressure coefficients. 

\subsubsection{Robustness to flap angle deflection ($\delta$)}
\label{ssec:delta}

\begin{figure}[h]
    \centering
    \includegraphics[width=\columnwidth]{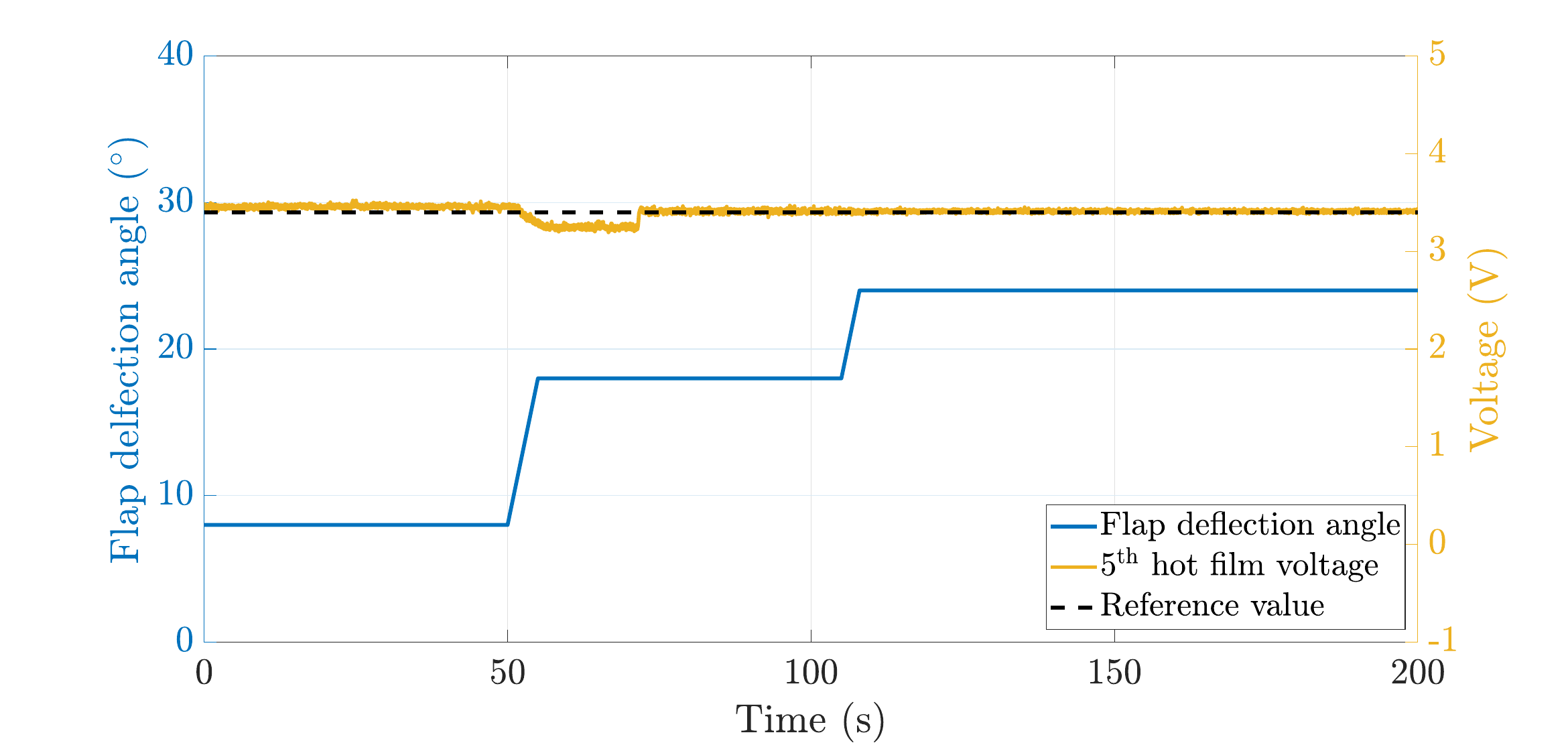}
    \caption{Evolution of the flap deflection angle against time (left axis) and evolution of both the $5 ^{\text{th}}$ hot film voltage and reference value (right axis) against time for the linear controller.}
    \label{fig:Control_Linear_Variation_Delta}
\end{figure}

In order to test the controllers robustness against the deflection angle, measurements were also performed with varying angles. In a given experiment, flap deflection angle was varied from $\delta = 8^\circ$ to $\delta = 18^\circ$ and to $\delta = 24^\circ$. The linear controller shows its limits due to its linear integral behaviour and the fact that it does not handle the positiveness of the system. As valves are not opened for $\delta =8^\circ$, the linear controller takes into account the error and therefore accumulates an integral error. In that sense, when the deflection angle increases to $\delta = 18^\circ$ for which valves have to be open, the controller effect is in the wrong direction. As observed in Figure \ref{fig:Control_Linear_Variation_Delta}, once the linear controller has overcome the accumulated error, the controller is robust to the deflection angle variation from $18^\circ$ to $24^\circ$. 

\begin{figure}[h]
    \centering
    \includegraphics[width=\columnwidth]{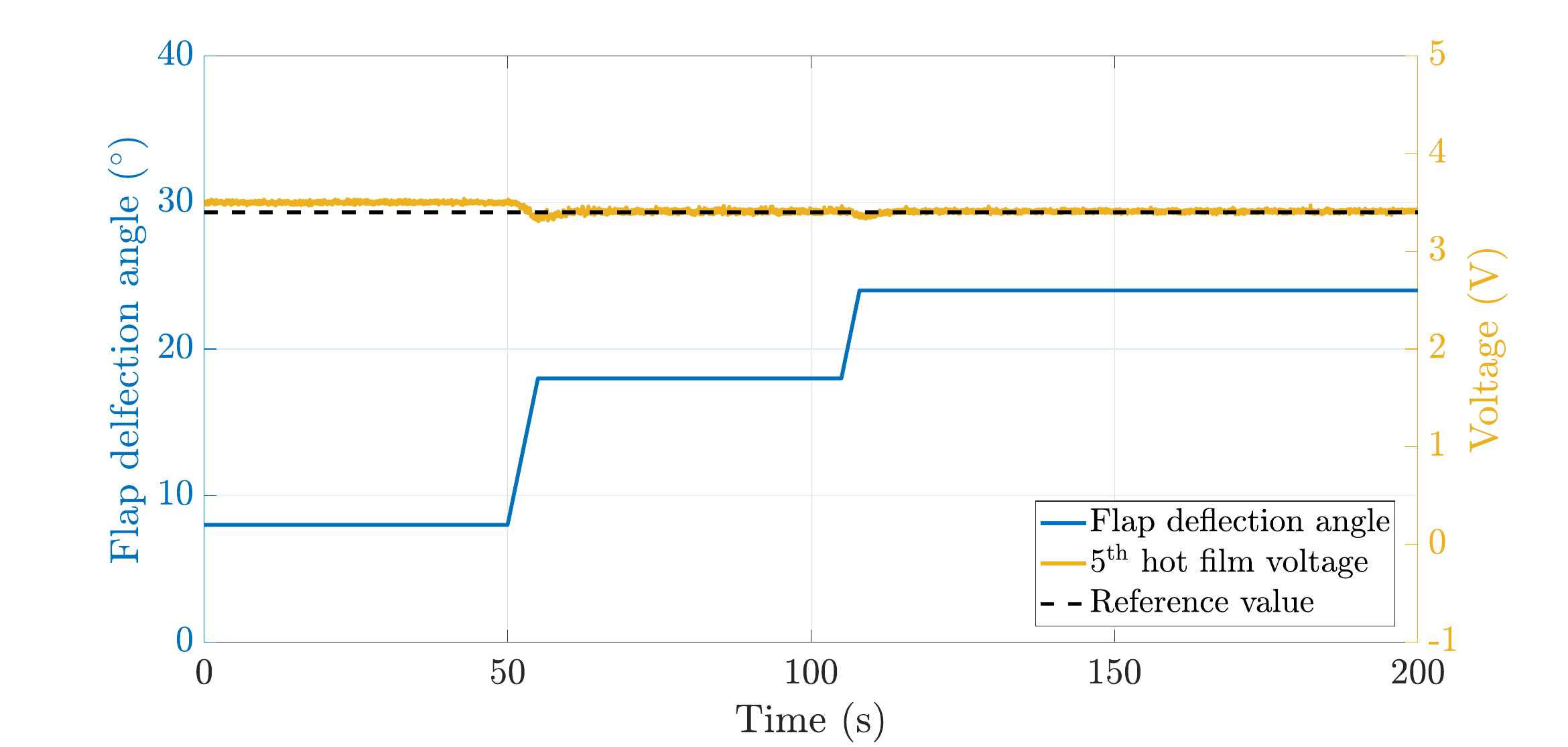}
    \caption{Evolution of the flap deflection angle against time (left axis) and evolution of both the $5^{\text{th}}$ hot film voltage and reference value (right axis) against time for the positive controller.}
    \label{fig:Control_Positive_Variation_Delta}
\end{figure}

As such an error could not be tolerated in real application, a way to circumvent it would be to implement an anti-windup on the linear controller. This may be done at a price of more involved calculus, usually involving a model, while here, the pure data-driven setup is employed. However, we proved in these experiments than an other simple and efficient way to deal with this integral behaviour is to implement a non-linear positive controller instead. As described in Figure \ref{fig:Control_Positive_Variation_Delta}, the delay resulting from the integral error issue totally vanishes considering this controller. In addition to that, in this setup, the input output stability is formally guaranteed.

\subsubsection{Frequency-domain responses}
\label{ssec:freq}

\begin{figure}[h]
    \centering
    \includegraphics[width=.8\columnwidth]{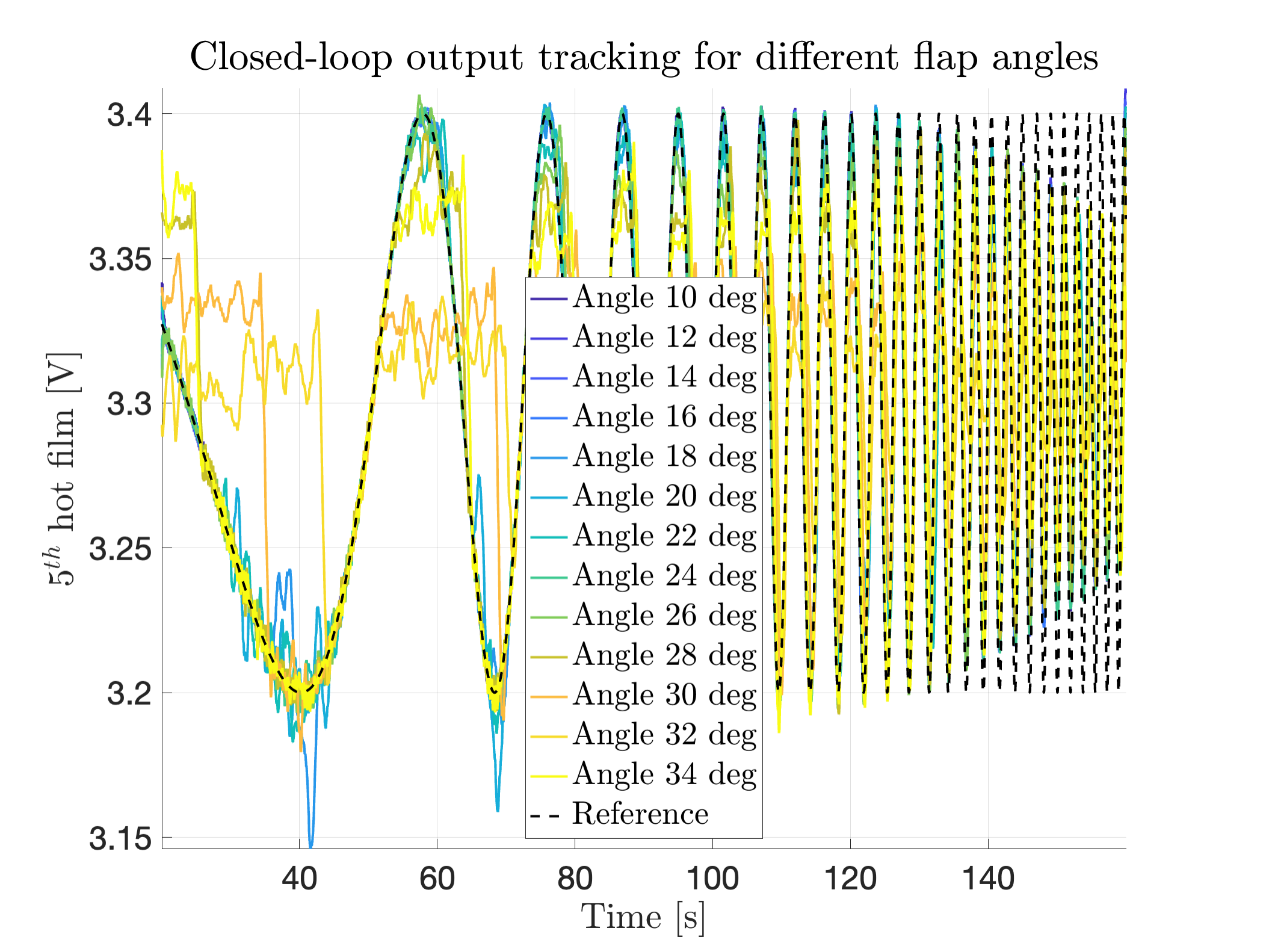}
    \includegraphics[width=.8\columnwidth]{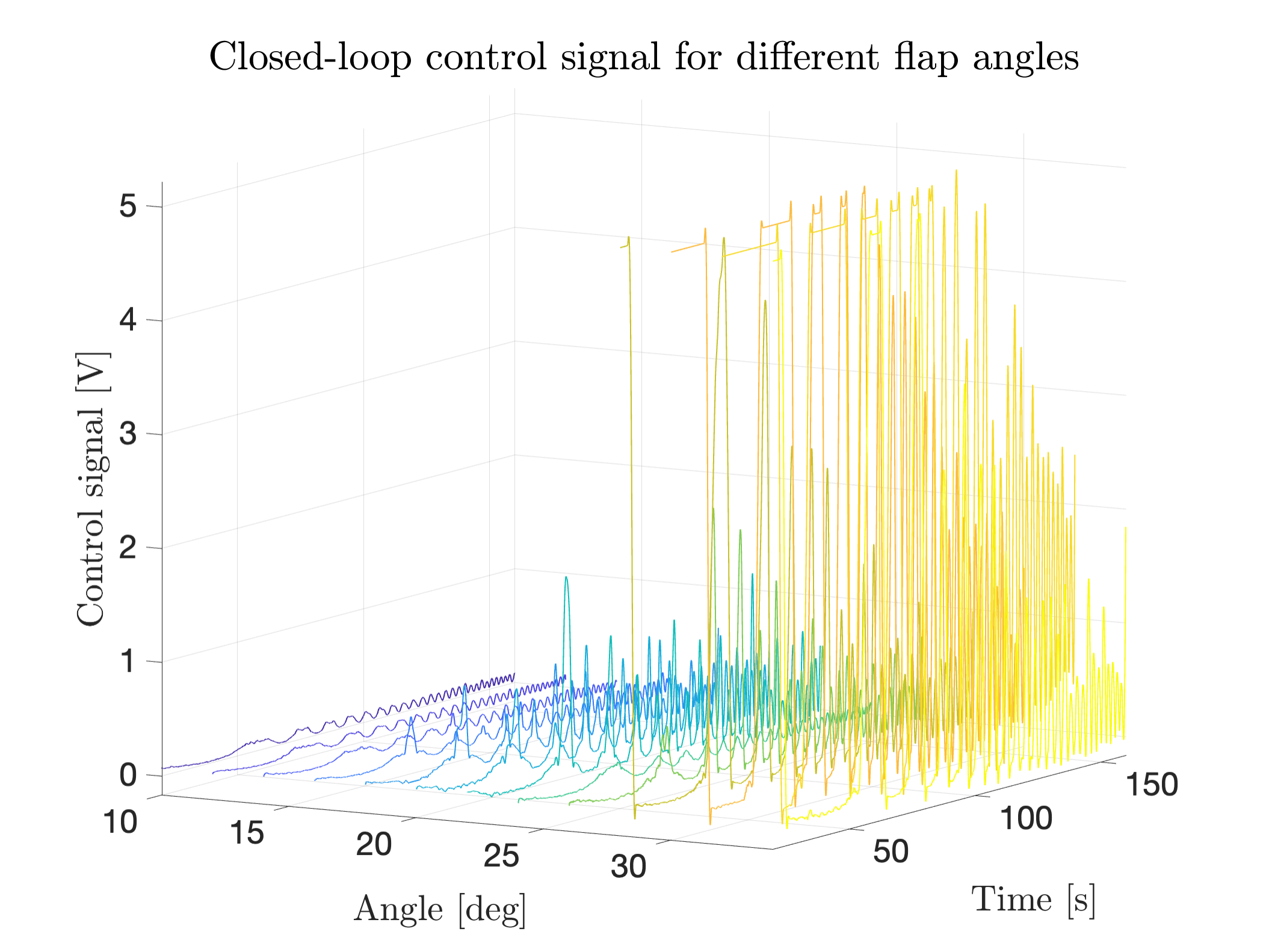}
    \caption{Linear controller action for frozen flap deflection angles $\delta$. Top: reference tracking performances. Bottom: produced control signals.}
    \label{fig:varyAngle_lin_sweep_y_u}
\end{figure}

In addition, to compute the frequency response of the closed-loop control and ensure that the closed-loop performances meet  the reference model $\mathbf M$ objective, a sinus around the tracking value, with frequency sweep signal is given as reference. The time-domain responses of the output and control signals are reported in Figure \ref{fig:varyAngle_lin_sweep_y_u}. 

The top frame allows computing the frequency response diagram given in the introduction, illustrating that the performances are very close to the expected one fixed by $\mathbf M$ (see Figure \ref{fig:performanceObjectiveFR}). On the bottom frame, the control signal of the linear controller shows to reach the saturation quite often, which is an other motivation for the nonlinear positive control.

\subsubsection{Further remark on the nonlinear positive control}
\label{ssec:positive}

\begin{figure}[h]
    \centering
    \includegraphics[width=.8\columnwidth]{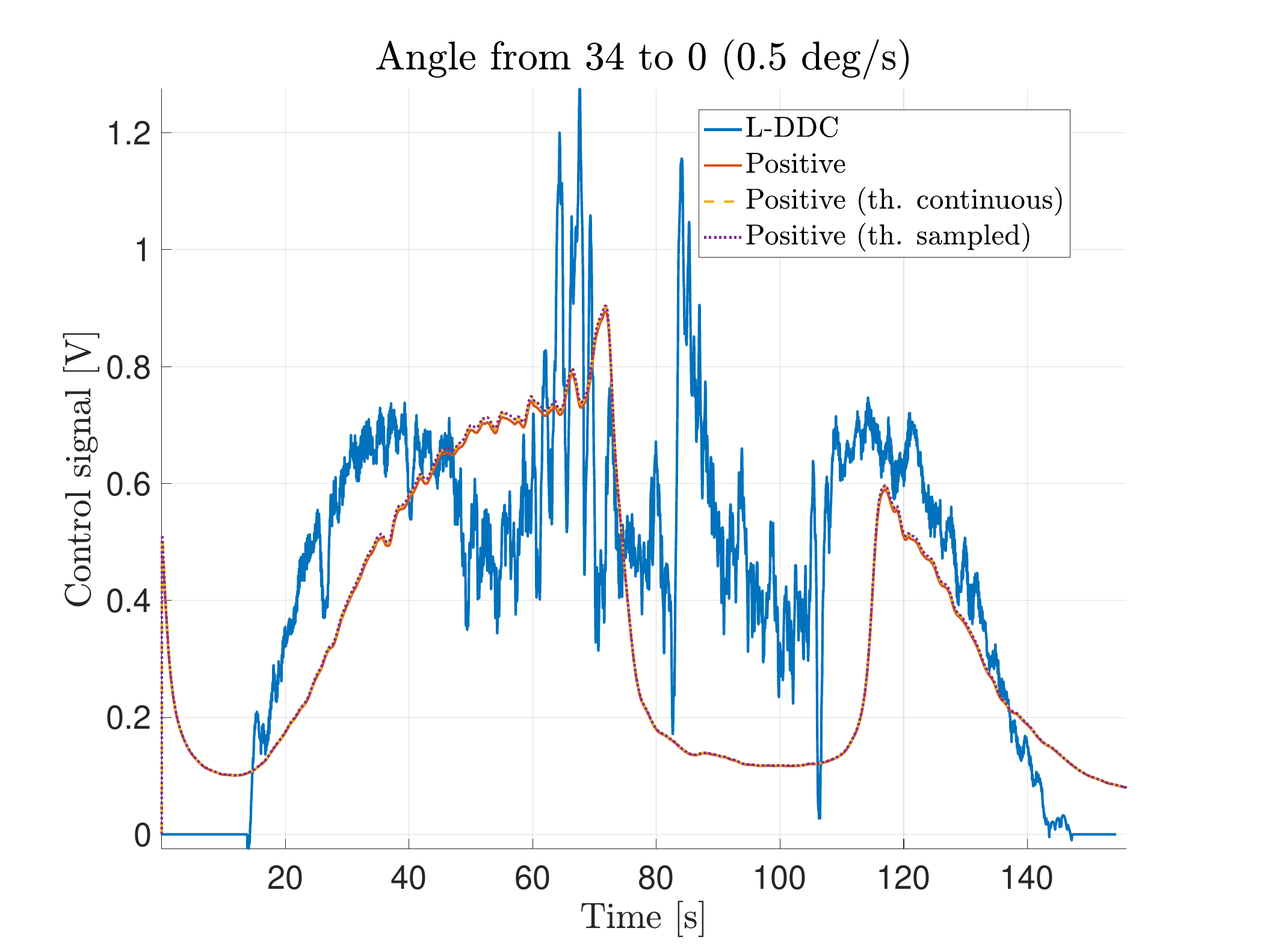}
    \includegraphics[width=.8\columnwidth]{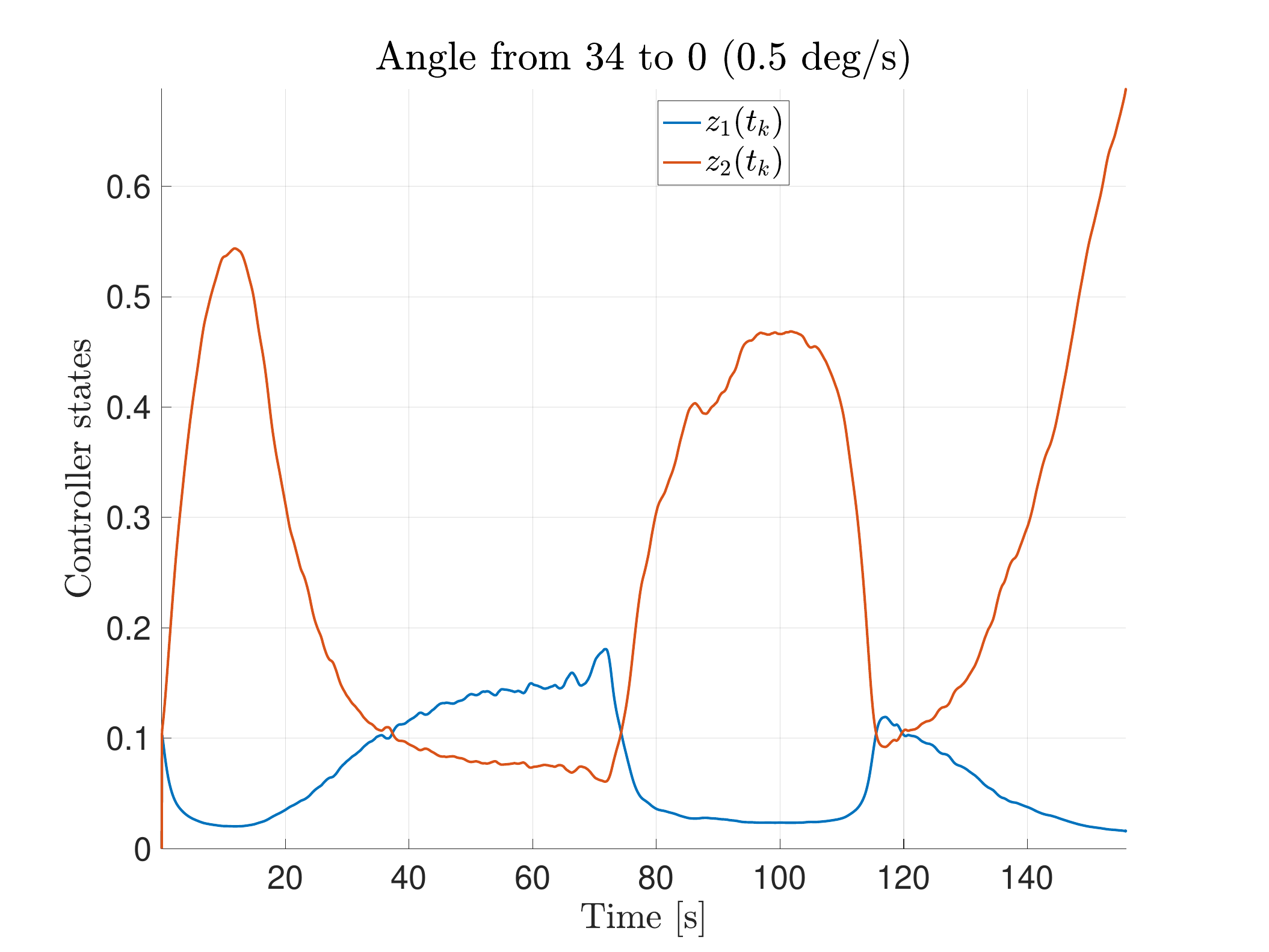}
    \caption{Experiment where flap angle $\delta$ travels from 34 degrees to 0 degree. Top: control signal of the linear and nonlinear experimental controllers and theoretical nonlinear continuous and sampled control. Bottom: sampled-time positive controller internal states}
    \label{fig:time_u}
\end{figure}

Finally, as the nonlinear positive control law seems to be the best solution, a time-domain experiment where the reference is fixed but the flap deflection angle $\delta$ travels from $34$ degrees to $0$ degree with a speed rate of $0.5$ deg/s is performed (we compare here both strategies). Figure \ref{fig:time_u} (top) shows the control signal actuation for both the linear and nonlinear (positive) controllers. First it illustrates the fact that the positive controller avoids the saturation while the linear one tends to often reach then. In the same frame, for the positive control law, we compare the experimental control signal (solid orange) with the reconstructed theoretical continuous \eqref{eq:positive_th} (dashed yellow) and sampled \eqref{eq:positive} (dashed violet) when fed by the experimental data. Both lead to a perfect match which confirms the good implementation. Then Figure \ref{fig:time_u} (bottom) illustrates the positive sampled-time controller internal states, which both remain, as expected, positives.

\section{Conclusions}
\label{sec:conclusion}

This paper presents experimental validations of active closed-loop control of flow separation over a plain flap. We numerically and experimentally demonstrate that flow separation may be improved by mean of a (SISO) reference signal tracking feedback control architecture, involving a controller with integral action. We also proposed two control laws: first \emph{(i)} a linear one where the integral gain is computed via a direct data-driven approach, and second, \emph{(ii)} a nonlinear positive controller to account for the system limitations, using the very same gain.

Both strategies enabled to maintain a reference voltage value on the objective hot film placed at the flap mid-chord. The latter reference, being calculated based on the flow separation observation. Both controllers efficiency were assessed through lift coefficient calculations derived from pressure measurements (Figure \ref{fig:performanceObjective}). Their robustness to the flap deflection angle was also tested through experiments, during which the flap angle was continuously varied. Additionally, we demonstrate that the expected theoretical performances were recovered experimentally. The most significant advantage of these control techniques lies in the simplicity of their application, which is of importance  for practitioners in view of experimentation and implementation. 



Application of such controllers could be extended to other flow control problems in future works, as well as more detailed validations.

\section*{Acknowledgments}
This work was funded by the French National Research Agency (ANR) in the framework of the ANR ASTRID MATURATION CAMELOTT-MATVAL Project. It is supported by the regional platform CONTRAERO in the framework of the CPER ELSAT 2020 Project. The Defence Innovation Agency (DIA) has also financially sustained this work. The authors also thank RENATECH, the French national nanofabrication network, and FEDER. It has also been financed by the ONERA research project FluiDyCon, Fluid Dynamical Control.

\bibliographystyle{plain}
\bibliography{_biblioPoussot,_biblioBriat,_biblioCPV,_biblio_TA}

\end{document}